# A comparison study of ridge filter parameter using FLUKA and GEANT4 simulation code

**Yongkeun Song, Jaeik Shin, Sungho Cho, Seunghoon Yoo**

**Ilsung Cho, Eunho Kim, Wongyun Jung***

*Division of heavy ion clinical research*

*Korea Institute of Radiological and Medical Sciences, Seoul 139-240*

**Sanghyoun Choi**

*Research center for radiotherapy*

*Korea Institute of Radiological and Medical Sciences, Seoul 139-240*

**Kyungmin Oh**

*Department of Biomedical Engineering*

*Inje University, Gimhae, Korea, 621-749*

We investigated the parameter optimization of ridge filter thickness using a Monte Carlo simulation for carbon ion therapy. For this study, a ridge filter was designed for the Spread-Out Bragg Peak (SOBP) by considering the relative biological effect (RBE). The thickness, height, and width of the ridge filter were designed by using the FLUKA and GEANT4 code, and we analyzed and compared the results of the physical dose distribution for the FLUKA and GEANT4 coding. The results show that the minimum width of the groove for the ridge filter should be at least 0.4cm for the appropriate biological dose. The SOBP sections are 8cm, 9cm, and 10cm, respectively, when heights are 3.5cm, 4.0cm, and 4.5cm. The height of the ridge filter is designed to be associated with the SOBP width. Also, the results for the FLUKA and GEANT4 code show that an average value of difference is 3% and a maximum error is 5%; however, its trend was similar. Therefore, the height and width of the groove for the ridge filter are used for important parameters to decide the length and plateau of SOBP.

Email: songyk0314@kirams.re.kr

Fax: +82-10-3863-4258

# I. INTRODUCTION

Carbon ion therapy is a very effective method for the treatment of cancer because of its very high Relative Biological Effectiveness (RBE) around the peak and excellent dose distribution. Carbon ion therapy was started in 1994 using the Heavy Ion Medical Accelerator in Chiba (HIMAC) at the National Institute of Radiological Sciences (NIRS) in Japan [1] since it has been researched for clinical trials on light ion therapy at Lawrence Berkeley National Laboratory, USA (LBNL) [2]. Heidelberg Ion Beam Therapy Center (HIT) in Germany also has been treating patient since 2009 [3]. Due to the high therapeutic efficiency [4], several countries plan to establish a treatment carbon ion therapy facility. Korea Institute of Radiological and Medical Sciences (KIRAMS) is currently under construction aiming clinical trial in 2017 at Busan Gijang in Korea.

The ion beam should be modulated for treating patients with an ion beam. One of methods for beam modulation is, broad beam, which is made to extend the Bragg Peak to the size of the tumor. Broad beam modulation method can be either active or passive. Active modulation method is used at GSI and passive modulation method is adapted at NIRS. In the passive modulation method, a special component should be added on the broad beam line; its ridge filter consist of a small groove. It is made up of aluminum fabricated in small steps and is used to create the pristine Bragg Peak to Spread Out Bragg Peak (SOBP). The design of these ridge filter has been developed by Kanai et al [5], Akagi et al [6] and Hata et al [7].

The SOBP is made by inserting bar ridge filters in the beam course [8]. In NIRS, the ridge filter is made of aluminum. The spacing of each bar ridge is 5 mm, and it does not move during irradiation. Due to multiple scattering in the ridge filter and the angular dispersion of the wobbled beam, shades of the bar ridge are smeared out at the irradiation site. In clinical trials, aluminum ridge filters are lined up for SOBP from 2-cm width to 12-cm width, in which case the height of the aluminum ridge becomes about 6 cm. For the SOBP of 15-cm width, a brass ridge filter is used [9], which can help select an appropriate ridge filter out of 8 ridge filters that are mounted on a large wheel. The ridge filter is designed so that the survival fraction of human salivary gland tumor cells (HSG cells) should be uniform in the SOBP [5].

In Korea Institute of Radiological and Medical Science (KIRAMS), a carbon therapy facility is under construction and both broad and scanning modes are being considered now. Thus, the ridge filter design is needed for developing a broad beam irradiation system. In this study, we analyzed the parameter effect on the dosimetric properties using MC simulation via the FLUKA and GEANT4 code. The comparison study of the ridge filter design was performed.

## II. MATERIAL

FLUKA (FLUktuierende KAskade) is a fully integrated Monte Carlo simulation package for the interaction and transport of particles and nuclei in matter [10, 11]. FLUKA has many applications in particle physics, high energy experimental physics and engineering, shielding, detector and telescope design, cosmic ray studies, dosimetry, medical physics, radiobiology. A recent line of development concerns hadron therapy.

In order to perform SOBP simulation of passive modulation, we designed with a ridge filter and water phantom in FLUKA 2011.2b code. The ridge filter is referred to Hyogo Ion Beam Center and Gunma Heavy Ion Medical Center of Gunma University (GHMC)'s paper [6, 12]. These papers compared the measured data with the Monte Carlo simulation results using the ridge filter by the carbon and proton beams, respectively.

The components were designed by the FLAIR geometry editor of the FLUKA additional tool. The groove of the ridge filter consists of a pair of the top and bottom wedge body as shown in figure 1(a). The spacing is inserted by between the grooves, and repeated same shape until distance 20cm as shown in figure 1(b). We also designed a thin aluminum plate, which has 0.3cm of thickness. Its purpose is hold up the groove. The height and width of grooves are important parameters to determine the length and slope of the SOBP. Due to the size of the tumor patient during treatment by replacing the different ridge filter to modulate the beam, the filter of various types should be designed and the results should be compared. In order to design 8cm SOBP of carbon ion, the height of the groove of the ridge filter should be designed by 3.8cm [12]. In the basement previous study, we designed a variation height and width of the grooves. The height length is from 3.0cm to 4.5cm, and the interval is 0.5cm. The width length is from 0.05cm to 0.7cm, and interval is 0.5cm. The spacing length is 0.1cm. We designed 56 ridge filter of different design. The ridge filter material is mostly used for aluminum. In this study, we assigned the material of ridge filter by density 2.699g/cm3 aluminum.

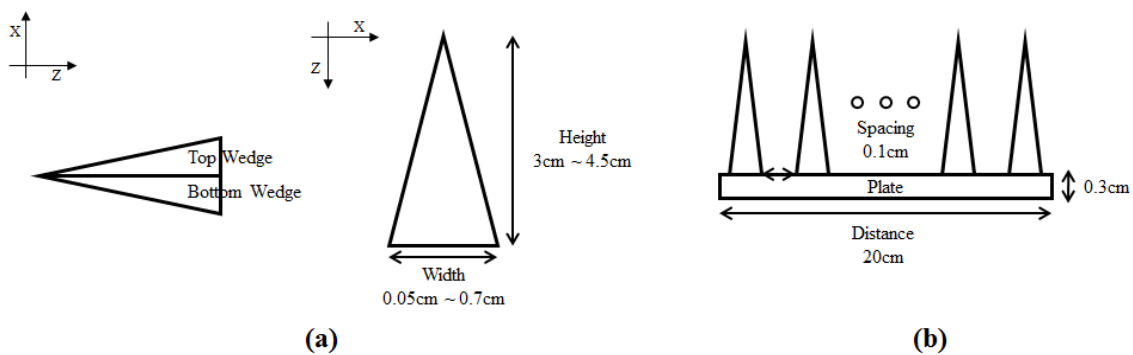

Figure 1. Schematic of groove design of ridge filter.

Figure 2 show geometry layout to perform the simulation of SOBP. The distance of between source and isocenter (SID) and source to ridge filter were 550cm and 420cm, respectively. The water phantom shape was square which dimension was 50 x 50 x 50cm. The distance of ridge filter to water phantom was 130cm. In addition, all of the components were included by air which shape was 2000 x 2000 x 2000cm cubic.

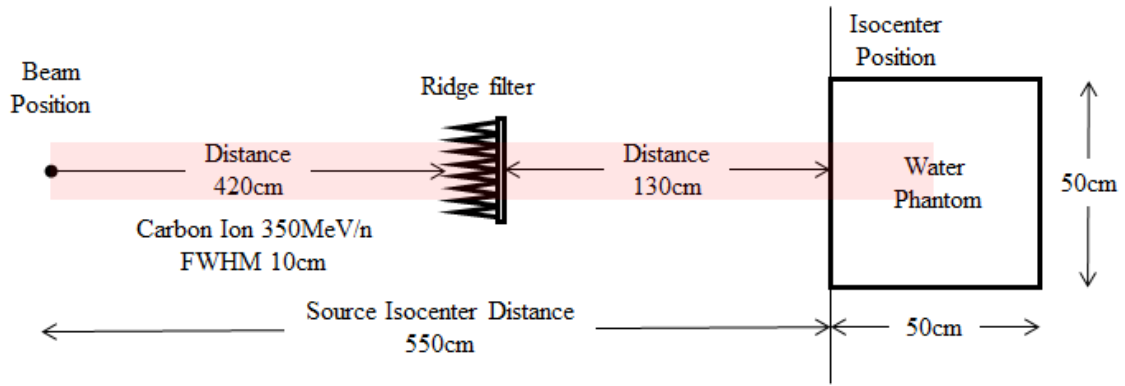

Figure 2. Schematic layout of broad beam delivery system in Monte Carlo simulation.

After geometry designed, in order to debug the geometry, we confirmed the components designed 2D and 3D as shown figure 3. Since FLUKA cannot run if find geometry error, it is essential to verify the region and body. The designed ridge filter was consisted of a lot of grooves because it is greater probability of error. We confirmed that there are no errors in the geometry, and went to next step.

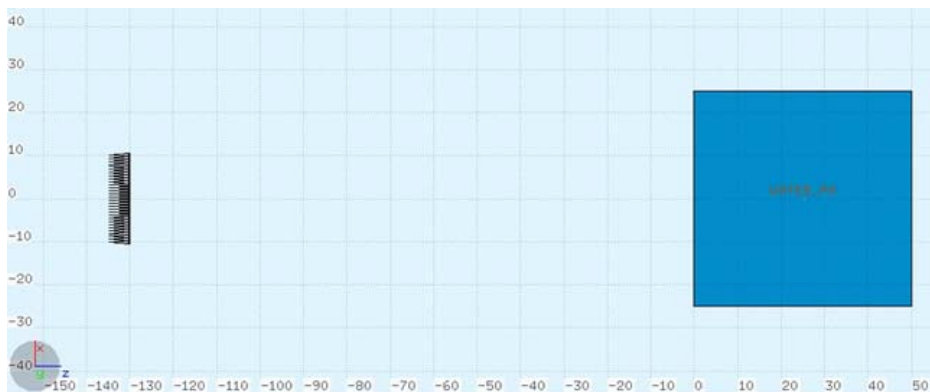

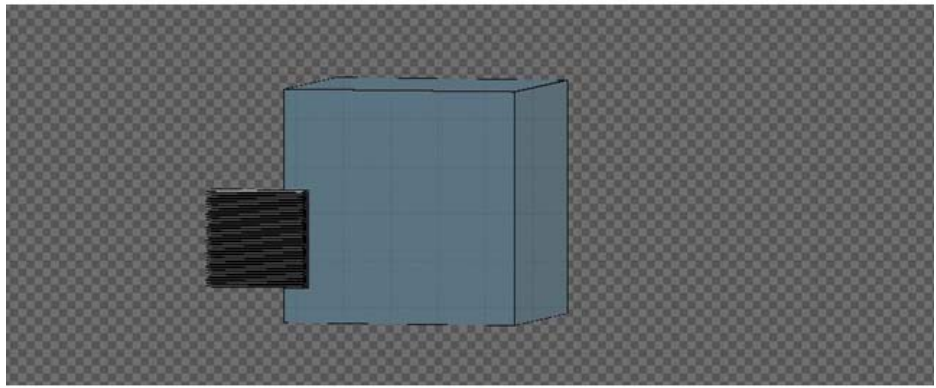

Figure 3. Geometry of 2D view (a) and 3D view (b) in FLUKA.

Beam parameter setup card in FLUKA are BEAM, HI-PROPE, BEAMPOS. Above mentioned these card can setup beam basic parameter such as beam energy, shape, divergence, initial position, and heavy ion type. We setup carbon beam using these card. Beam energy is 350MeV/n and FWHM (Full Width Half Maximum) of beam shape is 10cm, respectively. In addition, the distance between beam initial position and ridge filter is 420cm

Scoring used USRBIN card from among many scoring option in FLUKA. USRBIN card, which is a typical scoring card in FLUKA. This card checks the space distribution of the energy, or, is used in calculating the total fluence. The list of the scoring option of USRBIN includes deposited energy ($GeV/cm^3$), dose ($GeV/g$), activity ($Bq/cm^3$) and fluence (particles/$cm^2$). Since the main purpose of this study is to verify physical dose in water phantom, all of the scoring option setup the dose. In order to measure physical dose in water phantom, which with the same shape and the same size detector was added. To verify FWHM of initial beam when it penetrate the ridge filter, covered detector for ridge filter was added too. In addition, so as to reduce the voxel binning effect, scoring voxel size was setup 50cm x 50cm x 1cm.

Nuclear interactions generated by ions are treated through interfaces to external event generators. Heavy ion interaction model in FLUKA are Dual Parton Model (DPM), Relativistic Quantum Molecular Dynamics Model (RQMD) and Boltzmann Master Equation (BME). We have selected the RQMD for nucleus-nucleus interaction model.

In the last stage, we checked all of the card in FLUKA and determined number of initial particle. If number of initial particle increased, reduce the error. Thus, assigned too particles, it is inefficient on account of increased calculation time. Therefore, it is important to decide number of the initial particle. In this study, we carried out the simulation that initial particle number is $10^8$.

In order to verify results from FLUKA, we are also designed ridge filter and water phantom using GEANT4 among other Monte Carlo simulation code. Based on the geometry configuration information by designed using FLUKA, all of the parameters of components such as length of ridge filter and water phantom, source isocenter distance, material of components, and others are designed by as close as possible. In this study, we used the source codes and data libraries of GEANT4 10.0. The electromagnetic and hadron physics lists were based on a reference physics list. The electromagnetic physics process was set to "EM standard option 3" which is suitable for medical research. The hadron physics process was configured by reference physics list "QGSP_BIC_EMY", which included the Binary Cascade Model (BIC). Figure 4 indicates broad beam line layout and ridge filter in GEANT4, respectively. The beam is carbon ion 350MeV/n that FWHM is 10cm.

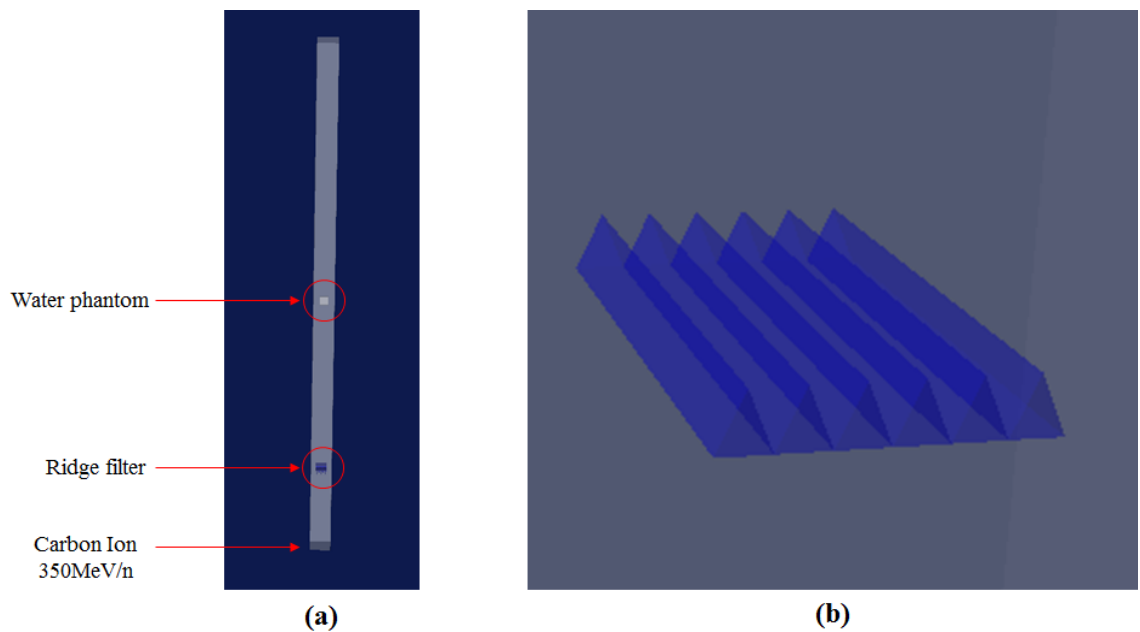

Figure 4. Geometry layout (a) and the ridge filter (b) in GEANT4.

## III. RESULTS

In order to ensure that the incident beam into ridge filter is correctly, we designed the detector on the ridge filter and measured absorbed dose. Figure 5(a) shows the 2D dose distribution on the ridge filter. It looks like 10cm x 10cm beam approximately. Since it is not accurate to verify beam Full

Width Half Maximum (FWHM), we converted 2D raw data to 1D profile owing to confirm the accuracy of beam. Figure 5(b) plots 1D beam profile on the ridge filter. It has confirmed that FWHM of beam is obtained as exactly 10cm.

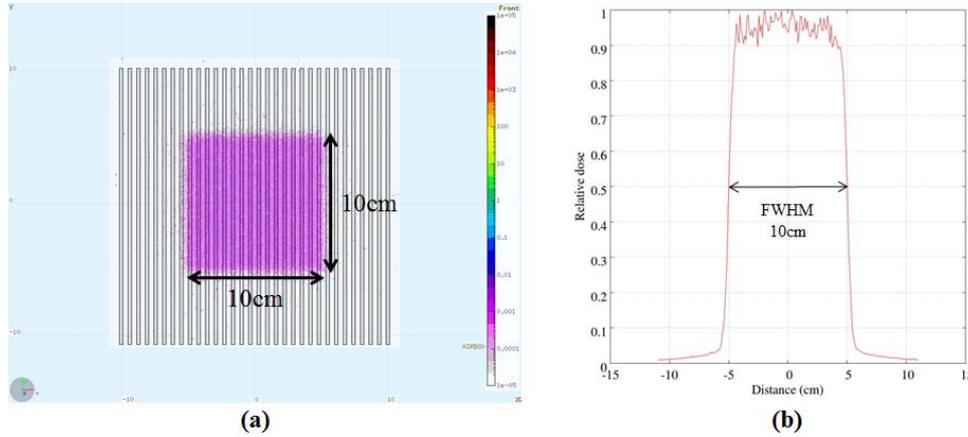

Figure 5. 2D dose distribution of the ridge filter (a) and 1D dose profile.

Figure 6 plots the depth distribution in water phantom for physical dose. We observed physical dose as changed that height length are 3.0cm, 3.5cm, 4.0cm, 4.5cm and width 0.05cm to 0.7cm and interval is 0.05cm, respectively. The each plot was normalized by a maximum value. When carbon ion beam pass through the ridge filter and incident water phantom, doses are gradually increased in until the beginning point of the SOBP. Then, it while having a constant slope is reduced to the beginning point of the tail. The length of width is below 0.4cm, second peak is occurred. Since carbon beam is not a sufficient attenuation below width 0.4cm, second peak occur at depth direction 21.5cm point of pristine Bragg Peak point of the carbon beam 350MeV/n in water phantom. In order to create a plateau section of the SOBP, the second peak should not occur. Therefore, we realize that the minimum width length of the groove of the ridge filter should be at least 0.4cm. Figure 6(b), 6(c) and 6(d) show length of height 3.5cm, 4.0cm, 4.5cm, respectively. The SOBP section of physical dose that length of height is 3.0cm, namely between the maximum dose and starting point at tail, observed 7cm of SOBP. When the length of height is 3.5cm, 4.0cm, and 4.5cm, SOBP section is 8cm, 9cm, and 10cm, respectively. We also realize that the length of height increased in 0.5cm whenever SOBP length also increases in about 1cm. Therefore, height and width of groove of ridge filter are used for an important parameter as length and plateau of SOBP.

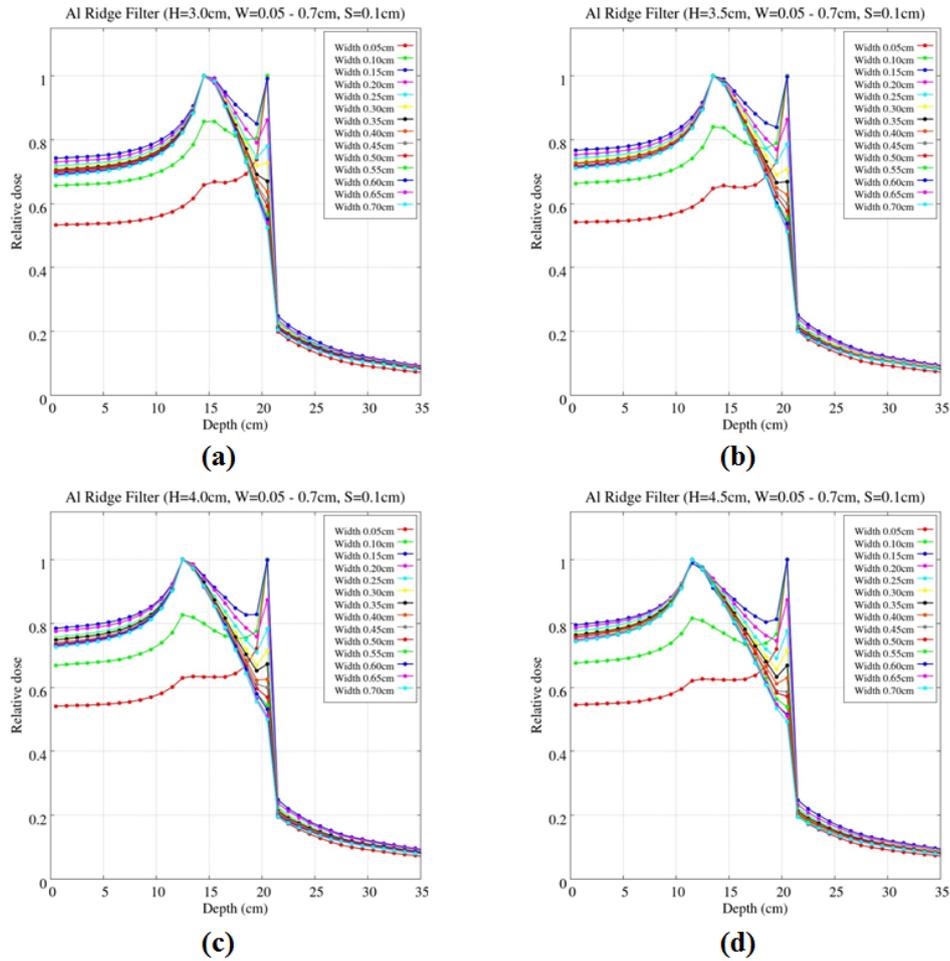

Figure 6. The physical dose distribution in a carbon ion beam of 350MeV/n was determined using FLUKA 2011 code. The heights of the ridge filters are 3.0cm (a), 3.5cm (b), 4.0cm (c) and 4.5cm (d), respectively. The width of the ridge filter is 0.05cm to 0.7cm and interval is 0.05cm.

In a previous study, we obtained physical dose regarding depth in water phantom using ridge filter of broad beam components by FLUKA 2011 code. In order to verify FLUKA results accuracy, we carried out other simulation code well-known GEANT4 code. And, FLUKA and GEANT4 results was compared. The all of the simulation setting such as geometry, beam, and detector are same compared FLUKA broad beam simulation. The number of the initial particle is $10^8$. As shown in figure 7, it compared with results of FLUKA and GEANT4. The results are height length of the ridge filter are 3.0cm, 3.8cm, and 4.5cm and width length are from 0.4cm to 0.7cm interval 0.1cm, respectively. The results showed that an average value of difference is 3% and a maximum error percentage is 5%. FLUKA and GEANT4 physical dose were compared by performing normalization as maximum dose.

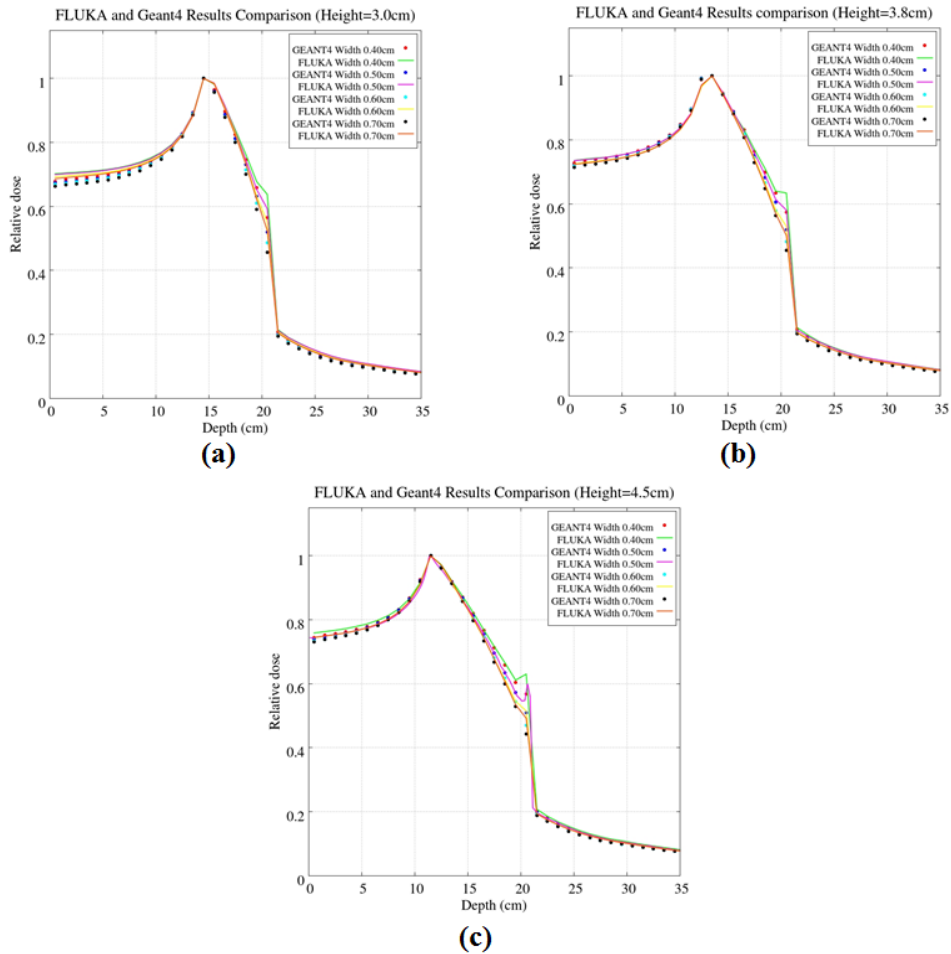

070

Figure 7. The calculation data of the FLUKA and GEANT4 is compared of the FLUKA and GEANT4 physical dose calculation. The height of the ridge filter are 3.0cm (a), 3.8cm (b) and 4.5 cm (c), respectively.

**IV. CONCLUSION**

In this study, we designed major component of broad beam which is the ridge filter using Monte Carlo simulation FLUKA and GEANT4. And then, the results of the physical dose distribution in water phantom were calculated. Furthermore, the height and width of the ridge filter changed value. Therefore, we realize the height and width as a parameter is to change the length and plateau of the SOBP, respectively. In addition, in order to determine the accuracy of the results, we compared FLUKA with GEANT4 results. It was confirmed that seen within the error of up to 5%. The ridge filter we designed should apply KHIMA project based on changed parameter, such as height, width and spacing. We plan to simulate Linear Energy Transfer (LET) and Relative Biological Effectiveness (RBE) using FLUKA. If we obtained LET and RBE, we would calculate biological dose.

## ACKNOWLEDGEMENT

This work was supported by the National Research Foundation of Korea (NRF) grant funded by the Korea government (MSIP) (201429534).

## REREFENCES